\documentclass[letter,twocolumn]{jpsj2}
\def\runauthor{\textsc{S.~Watanabe} and \textsc{M.~Imada}}
\makeatletter
\def\@typeset{}
\makeatother
\def\gsim{\mathop {\vtop {\ialign {##\crcr 
$\hfil \displaystyle {>}\hfil $\crcr \noalign {\kern1pt \nointerlineskip } 
$\,\sim$ \crcr \noalign {\kern1pt}}}}\limits}
\def\lsim{\mathop {\vtop {\ialign {##\crcr 
$\hfil \displaystyle {<}\hfil $\crcr \noalign {\kern1pt \nointerlineskip } 
$\,\,\sim$ \crcr \noalign {\kern1pt}}}}\limits}

\title{
On Proximity of 4/7 Solid Phase of $^3$He Adsorbed on Graphite \\
-Origin of Specific-Heat Anomalies in Hole-Doped Density-Ordered Solid-
}

\author{Shinji \textsc{Watanabe} and Masatoshi \textsc{Imada}}

\inst{Department of Applied Physics, University of Tokyo, Hongo 7-3-1, 
Bunkyo-ku, Tokyo, 113-8656, Japan}

\abst{
We theoretically study the stability of the solidified 2nd-layer $^3$He at 4/7 
of the 1st-layer density adsorbed on graphite, which exhibits quantum spin liquid. 
We construct a lattice model for the 2nd-layer $^3$He 
by taking account of density fluctuations on the 3rd layer 
together by employing 
the refined configuration recently found by path integral Monte Carlo simulations. 
When holes are doped into the 4/7 solid, within the mean-field approximation, 
the density-ordered fluid emerges. 
The evolution of hole pockets offers a unified explanation for the measured 
doping and temperature dependences of 
specific-heat anomalies. 
We argue that 
differentiation in momentum space is a key to understanding the physics and 
accounts for 
multiscale thermodynamic anomalies in the mono- and double-layered $^3$He systems 
beyond the mean-field level. 
}

\kword{quantum spin liquid, $^3$He, 4/7 phase, density-ordered fluid, hole pocket, zero point vacancy}

\begin{document}
\maketitle


Layered $^3$He systems adsorbed on graphite offer unique playgrounds 
as an ideal prototype of strongly correlated Fermion systems 
in purely two dimensions. 
At the 4/7 density of the 2nd-layer $^3$He relative to the 1st-layer density, 
$^3$He atoms are solidified to form a triangular lattice, which is referred to as the 4/7 phase~\cite{Elser}. 
The 4/7 phase has attracted much attention recently, since specific heat~\cite{ishida} 
and susceptibility measurements down to $10~\mu$K~\cite{masutomi} have suggested the 
emergence of a gapless quantum spin liquid. 

Theoretically, the multiple-spin-exchange model~\cite{Roger} has been used 
for the analysis~\cite{misguich,momoi}, which, however, has left fundamental questions about the 4/7 phase: 
(1) How to realize the gapless ground state? 
(2) How to explain the large saturation field of about 10~T~\cite{ishimoto}?
To resolve these issues, we have pointed out the importance of density fluctuations 
between the 2nd and 3rd layers~\cite{WI2007}. 
By constructing a lattice model to describe the 4/7 solid phase, 
we have shown that strong density fluctuations indeed stabilize a gapless quantum spin liquid, 
which can be regarded as that caused 
essentially by the same mechanism found 
in the Hubbard model~\cite{KI,MWI,Watanabe,Mizusaki}.
Furthermore, the density fluctuation accounts for the 
enhanced saturation field as observed~\cite{WI2007}. 

In this letter, we report our analysis of the stability of the 4/7 phase on the basis of the lattice model, 
taking account of density fluctuations. 
The model is derived from a refined configuration of the 2nd-layer $^3$He 
recently revealed by path-integral Monte Carlo (PIMC) simulations~\cite{takagi}. 
Our mean-field (MF) results 
show that a density-ordered fluid emerges when holes are doped into the 4/7 phase and 
that the evolution of the hole pockets explains measured specific-heat anomalies. 
Our model gives a unified explanation 
of unusual temperature and doping dependences of specific heat over a range 
from the 4/7 solid phase to the uniform-fluid phase at low densities. 
We also discuss the validity of the present picture beyond the MF approximation 
and possible relevance to the specific-heat anomalies measured recently 
in the double-layered $^3$He system~\cite{saunders}. 

Our analysis starts from the result of  
recent PIMC simulations~\cite{takagi},
which have revealed a more stable configuration 
for the 4/7 phase than that first proposed by Elser~\cite{Elser,PIMC} as shown in Fig.~1(a). 
Here, the open circles represent the atoms on the 1st layer, and the shaded circles, the locations 
on the 2nd layer when solidified. 
If $^3$He ($^4$He) atoms are adsorbed on the 1st layer, it forms a triangular lattice 
with the lattice constant $a$=3.1826 (3.1020)~\AA~at the saturation density 
$\rho_1$=0.114 (0.120)~atom/\AA$^2$~\cite{Elser}. 

\begin{figure}[h]
\begin{center}
\includegraphics[width=7.2cm]{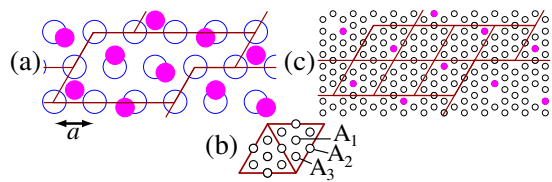}
\end{center}
\caption{
(Color online) 
(a) Lattice structure of the 4/7 phase of $^3$He shown by recent PIMC simulations~\cite{takagi}. 
Both 1st-layer atoms (open circles) and 2nd-layer atoms (shaded circles) 
form triangular lattices in the solid phase. 
The area enclosed by the solid line represents the unit cell for the solid 
of the 2nd-layer atoms (see text). 
The lattice constant of the 1st layer is $a$. 
(b) Possible stable location of the 2nd-layer atoms are shown by circles 
on top of a $a \times a$ parallelogram constructed from the 4 neighboring 1st-layer atoms.
(c) Structure of discretized lattice for the 2nd-layer model. Lattice points are shown by circles.
}
\label{fig:lattice77}
\end{figure}

The location of the 2nd-layer atoms is in principle determined as stable points in continuum space 
under the periodic potential of the 1st layer.  
In the present treatment, we simplify the continuum by discretizing it with the largest possible number of 
lattice points kept as candidates of stable points in the solid.
To illustrate discretization, we cut out from Fig.~1(a) 
a unit cell of the 1st layer, namely, a parallelogram 
whose corners are 
the locations on top of the neighboring 
4 atoms on the 1st layer as in Fig.~1(b). 
Possible stable locations of $^3$He atoms on the 2nd layer 
are points on top of (1) the centers of the 1st-layer triangle (A$_1$), 
(2) the midpoint of two-neighboring 1st-layer atoms (A$_2$), 
and (3) a nearby point of the 1st-layer atom (A$_3$). 
Therefore, we employ totally 11 points as discretized lattice points
in a parallelogram, shown as circles in Fig.~1(b). 
Since a unit cell in Fig.~1(a) contains 7 parallelograms, 
it contains 77 lattice points in total, as illustrated as circles in Fig.~1(c). 
Now the 4/7 solid phase is regarded as a regular alignment of 4 atoms on 77 available lattice points 
in the unit cell shown in Fig.~1(c).

For inter-helium interaction, 
we employ the Lennard-Jones potential 
$
V_{\rm LJ}(r)=4\epsilon \left[
\left(
\sigma/r
\right)^{12}
-
\left(
\sigma/r
\right)^{6}
\right]$, 
where $\epsilon=10.2$ K and $\sigma=2.56$ \AA~\cite{Boer}.
A more refined Aziz potential is expected to give similar results under this discretization. 
The interaction between $^3$He atoms on the 2nd layer 
is given by $H_{V}=\sum_{ij}V_{ij}n_i n_j$ 
($n_i$ is the number operator of a Fermion on the $i$-th site) 
with $V_{ij}$ taken from the spatial dependence of 
$V_{\rm LJ}(r)$ 
on the lattice points. 
In the actual $^3$He system, the chemical potential of the 3rd layer is estimated to be 16 K higher than 
that of the 2nd layer~\cite{Whitlock}. 
$^3$He atoms may fluctuate into the 3rd layer over this chemical potential difference and  
it is signaled by a peak of the specific heat at $T\sim 1$ K~\cite{Sciver,Graywall,Fukuyama2007}. 
To take account of this density fluctuation, we here mimic the allowed occupation on the 3rd layer by introducing a simple finite cutoff $V_{\rm cutoff}$  
for $V_{ij}$ within the same form of Hamiltonian: When $V_{\rm LJ}(r_{ij})$ for $r_{ij}\equiv |{\bf r}_i-{\bf r}_j|$ exceeds $V_{\rm cutoff}$, we take $V_{ij}=V_{\rm cutoff}$, 
otherwise $V_{ij}=V_{\rm LJ}(r_{ij})$. This allows us to take account of the qualitative but essential part of the possible occupation on the 3rd layer by atoms overcoming $V_{\rm cutoff}$. 

We consider the spinless-Fermion model on the lattice 
$H=H_{\rm K}+H_{\rm V}$ where the kinetic energy is given by 
$H_{\rm K}=-\sum_{\langle ij\rangle}(t_{ij}c_{i}^{\dagger}c_j 
+{\rm H.C.})$. 
We ignore the effects of corrugation potential from the 1st layer for the moment and 
discuss it later. 
By using the unit-cell index $s$ and the site index $l$ in the unit cell, 
we have ${\bf r}_i={\bf r}_s+{\bf r}_l$. 
After the Fourier transform, 
$c_{i}=c_{s,l}=\sum_{\bf k}c_{{\bf k},l}e^{{\rm i}{\bf k}\cdot{\bf r}_s}/\sqrt{N_{\rm u}}$, 
the MF approximation with the diagonal order parameter 
$\langle n_{{\bf k},l}\rangle$ leads to
\begin{eqnarray}
H_{\rm V}\sim H_{\rm V}^{\rm MF}
\mbox{\qquad\qquad\qquad\qquad\qquad\qquad\qquad\qquad\qquad\qquad}
\nonumber \\
=
\frac{1}{N_{\rm u}}
\sum_{l,m=1}^{77}
\sum_{s'}V^{lm}(s')
\sum_{{\bf k}, {\bf p}}
\left[
\langle n_{{\bf k},l} \rangle n_{{\bf p},m}
-
\frac{1}{2}
\langle n_{{\bf k},l} \rangle \langle n_{{\bf p},m}\rangle
\right], 
\nonumber
\end{eqnarray}
where the inter-atom interaction is expressed as 
$V_{ij}=V_{st}^{lm}=V^{lm}(s')$ 
with ${\bf r}_{s'}={\bf r}_{s}-{\bf r}_{t}$. 
Then, we have the MF Hamiltonian 
$H_{\rm MF}=H_{\rm K}+H_{\rm V}^{\rm MF}$. 
By diagonalizing the $77 \times 77$ Hamiltonian matrix 
for each ${\bf k}$, 
we obtain the energy bands 
$
H_{\rm MF}=\sum_{\bf k}\sum_{l=1}^{77}
E_{l}({\bf k})c_{{\bf k},l}^{\dagger}c_{{\bf k},l}. 
$

For the kinetic energy, several choices of $t_{ij}$ are examined. 
As noted in ref.~\citen{WI2007}, 
if $t_{0}$ is determined so as to reproduce the total kinetic energy $E_{\rm K}^{\rm PIMC}$, 
the main result below, measured in the unit of K, is quite insensitive to the choice 
of $t_{ij}$. 
Hence, we here show the results for $t_{ij}=t_{0}$ 
for the $ij$ pairs up to the shortest-19th 
$r_{ij}$, i.e., for $|r_{ij}|\le 2a$. 
The interaction $V_{ij}$ is taken for $r_{ij}\le 2a$, since the longer $r_{ij}$ part 
is ineffective~\cite{WI2007}. 
The recent PIMC simulation estimates the kinetic energy of the 2nd-layer 
$^3$He in the 4/7 phase as $E_{\rm K}^{\rm PIMC}=14$~K (17~K) for the $^4$He ($^3$He) 
1st layer~\cite{takagi}. 
Thus, we evaluate $t_0$ by imposing the condition, 
$
\sum_{\langle ij \rangle}t_{0}\langle c_{i}^{\dagger}c_{j}+{\rm H.C.}\rangle 
/(4N_{\rm u})=E_{\rm K}^{\rm PIMC} 
$
as $t_0=413$~mK (502~mK) for the $^4$He ($^3$He) 1st layer.

\begin{figure}
\begin{center}
\includegraphics[width=7.5cm]{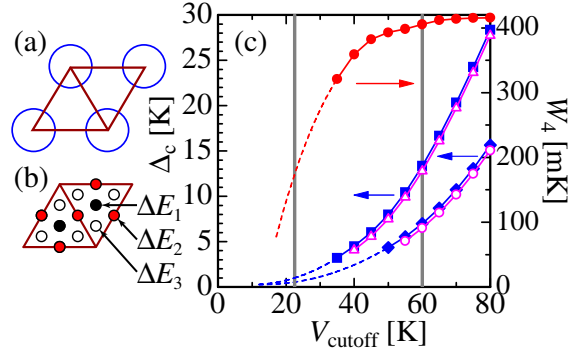} 
\end{center}
\caption{
(Color online) 
(a) Parallelogram formed by 4-neighbouring 1st-layer atoms. 
(b) The sites for 2nd-layer atoms in the parallelogram under 
corrugation potentials 
$\Delta E_1$ (solid circle), $\Delta E_2$ (shaded circle) and $\Delta E_3$ (open circle). 
(c) $V_{\rm cutoff}$ dependence of the ``charge gap'' $\Delta_{\rm c}$ 
for $\Delta E_1$=$\Delta E_2$=$\Delta E_3$=0 (filled square) and for $\Delta E_1=-3.0$~K, 
$\Delta E_2=-1.5$~K 
and $\Delta E_3=0$~K (open triangle) in the $^3$He/$^4$He/Gr system, and 
for $\Delta E_1$=$\Delta E_2$=$\Delta E_3$=0 (filled diamond) and for $\Delta E_1=-3.0$~K, $\Delta E_2=-1.5$~K 
and $\Delta E_3=0$~K (open circle) in the $^3$He/$^3$He/Gr system. 
The bandwidth of the 4th band $W_4$ is illustrated 
for $\Delta E_1$=$\Delta E_2$=$\Delta E_3$=0 (filled circle) 
in the $^3$He/$^4$He/Gr system. 
Shaded lines at $V_{\rm cutoff}=22.5$~K and 60~K are guides for the eyes (see text). 
}
\label{fig:cgp}
\end{figure}

By solving the MF equations for $H_{\rm MF}$, we have the solution of the 
$\sqrt{7}\times \sqrt{7}$ commensurate structure shown in Fig.~\ref{fig:lattice77}(c) 
with opening the ``charge gap" as shown by filled squares (diamonds) in Fig.~\ref{fig:cgp}(c) 
when $^4$He ($^3$He) is adsorbed on the 1st layer. 
Here, the ``charge gap" is defined by 
$\Delta_{\rm c}\equiv E_5^{\rm min}({\bf k})-E_4^{\rm max}({\bf k})$, where 
$E_l^{\beta}({\bf k})$ denotes the minimum or maximum energy of the $l$-th band 
from the lowest. 
We show here the well-converged results for large $\Delta_{\rm c}$ 
and the small-$\Delta_{\rm c}$ region will be discussed later 
for detailed comparison with experiments. 
We now discuss the effects of corrugation potential. 
PIMC~\cite{PIMC,takagi} suggests that 
the 1st-layer atoms make the corrugation potential on the 2nd layer even within 
the discretized lattice points. 
It suggests $\Delta E_1=-3.0$~K and $\Delta E_2=-1.5$~K relative to $\Delta E_3$ in the notation 
in Fig.~\ref{fig:cgp}(b). 
This effect 
merely shifts the $\Delta_{\rm c}$-$V_{\rm cutoff}$ line toward a larger $V_{\rm cutoff}$, 
as shown by open triangles (circles) for 
the $^4$He ($^3$He) 1st layer in Fig.~\ref{fig:cgp}(c).
Hence, we show our results below for $\Delta E_1=\Delta E_2=\Delta E_3=0$ and 
for $^4$He adsorbed on the 1st layer as a representative case~\cite{memo_dE}. 
We note that Fig.~\ref{fig:cgp}(c) indicates that the 4/7 phase is 
more stable when $^4$He is adsorbed on the 1st layer rather than the $^3$He 1st layer. 

\begin{figure}
\begin{center}
\includegraphics[width=5.5cm]{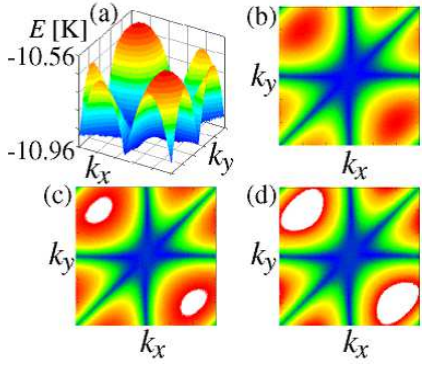} 
\end{center}
\caption{
(Color) (a) The 4th energy band in the folded Brillouin zone for 
$-\pi/\bar{a}\le k_x \le \pi/\bar{a}$ 
and $-\pi/\bar{a}\le k_y \le \pi/\bar{a}$ 
with $\bar{a}=2\sqrt{7}a$~\cite{memoBZ} 
for $V_{\rm cutoff}=60$~K at $n_0=1$. 
Contour plot of the 4th band for (b) $n_0=1$, (c) $n_0=0.99$ and (d) $n_0=0.97$. 
Hole pockets are represented by white regions in (c) and (d). 
}
\label{fig:FermiSF}
\end{figure}

The energy band $E_{4}({\bf k})$  for $V_{\rm cutoff}=60$~K at $n_{0}=1$ with 
$n\equiv \sum_{l=1}^{77}\sum_{\bf k}n_{{\bf k},l}/(77 N_{\rm u})=4n_{0}/77$ 
and its contour plot in the folded Brillouin zone~\cite{memoBZ} are shown 
in Figs.~\ref{fig:FermiSF}(a) and \ref{fig:FermiSF}(b), respectively. 
When holes are doped into the 4/7 phase, 
in our MF results, 
the Fermi surface appears at the 4th band 
as hole pockets for $n_{0}=0.99$ 
(Fig.~\ref{fig:FermiSF}(c)) and $n_{0}=0.97$ (Fig.~\ref{fig:FermiSF}(d)) 
with the density order retained. 

\begin{figure}
\begin{center}
\includegraphics[width=8.7cm]{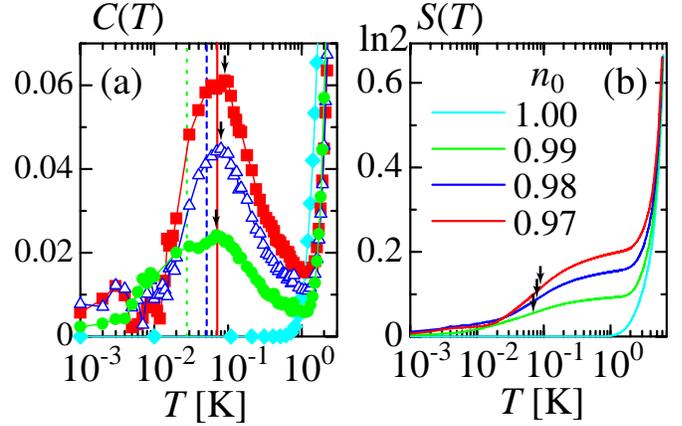} 
\end{center}
\caption{
(Color) (a) Specific heats for $n_0=1.0$ (light-blue diamond), 0.99 (green circle), 
0.98 (blue open triangle), and 0.97 (red square) for $V_{\rm cutoff}=60$~K. 
$W_4^{\rm top}-\mu$ for $n_0=0.99$ (green dotted line), 0.98 (blue dashed line), 
and 0.97 (red line). 
(b) Entropy per site for $n_0=1.0$ (light blue), 0.99 (green), 0.98 (blue), 
and 0.97 (red) for $V_{\rm cutoff}=60$~K. 
Arrows indicate the temperatures $T^{*}$ at which $C(T)$ has the highest peak 
for $T<T_{\rm c}$. 
}
\label{fig:CTST}
\end{figure}

This evolution of hole pockets is reflected in thermodynamic quantities: 
As holes are doped into the 4/7 solid, a remarkable peak in the specific heat 
$C(T)$ develops at temperatures lower than 
the density-order transition temperature $T_{\rm c}$, as shown in Fig.~\ref{fig:CTST}(a). 
At the same time, $C(T)$ at temperatures right below $T_{\rm c}$ 
decreases. 
The temperature $T^{*}$ at which $C(T)$ has the highest peak for $T<T_{\rm c}$ 
increases as $n_0$ decreases from 1, as indicated by arrows in Fig.~\ref{fig:CTST}(a). 
This tendency was actually observed in the past~\cite{Sciver} and 
recent~\cite{Fukuyama2007} measurements: 
The $\lambda$-like anomaly at $T=T_{\rm c}\sim 1$~K simultaneously with 
the hump at $T\sim 40$~mK in $C(T)$ observed for $n_{0}=0.96$~\cite{Sciver} and 
for $n_{0}=0.971$~\cite{Fukuyama2007} 
is an indication of the coexisting density order and fluid. 
Since our model $H$ is the spinless-Fermion model, the entropy per site $S(T)$ 
at the high-temperature limit is given by $k_{\rm B}$ln2. 
When $T$ decreases, $S(T)$ sharply drops at the density order $T_{\rm c}$ and the 
remaining entropy is released at $T\sim T^{*}<T_{\rm c}$ as shown in Fig.~\ref{fig:CTST}(b). 

In experiments, sharp decreases in $S(T)$ at much lower temperatures, i.e., 
$T\sim 0.2$ and $1$~mK, 
are observed for $0.9\lsim n_{0} \lsim 1.02$~\cite{ishida,Fukuyama2007}, 
which result in a double peak in $C(T)$. 
This spin entropy is ignored in the present model. 
These sharp drops correspond to the 
release of the spin entropy at the energy scale of the 
spin exchange interaction $J$~\cite{WKB}. 
Under hole doping, the double peak 
at $T\sim 0.2$ and 1~mK is suppressed, whereas 
the hump $C(T^{*})$~\cite{Fukuyama2007} at 40~mK grows. 
This is naturally understood by the 
suppression of the density order 
simultaneously with an 
increase in the fluid contributions induced 
by doping. 
Actually, a double-peak structure in $C(T)$ around $T\sim J$ 
has been shown by our exact-diagonalization calculations~\cite{WI2007} for the density-ordered solid phase 
in a minimal lattice model, which was introduced to mimic the 4/7 solid. 

\begin{figure}
\begin{center}
\includegraphics[width=7.5cm]{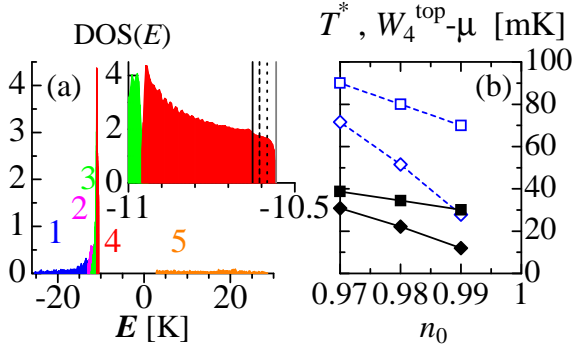} 
\end{center}
\caption{
(Color) (a) Density of states for the 1st (blue), 2nd (pink), 
3rd (green), 4th (red), and 5th (light brown) bands for $V_{\rm cutoff}=60$~K at $n_0=1$. 
The inset shows the enlargement of the 4th band 
with $W_4^{\rm top}$ (gray line) and the chemical potential $\mu$ 
for $n_{0}=0.99$ (dotted line), 0.98 (dashed line), and 0.97 (solid line). 
(b) Filling $n_0$ dependence of $T^{*}$ (square) and $W_4^{\rm top}-\mu$ (diamond) 
for $V_{\rm cutoff}=60$~K (open symbols with dashed lines) and 
$V_{\rm cutoff}=22.5$~K (filled symbols with solid lines). 
}
\label{fig:DOS_Tstar}
\end{figure}

To further clarify the origin of $T^{*}$, we calculated the density of states 
for $V_{\rm cutoff}=60$~K at $n_{0}=1$, as in Fig.~\ref{fig:DOS_Tstar}(a). 
Since the energy gap opens in the 4/7 phase, $\Delta_{\rm c}\sim 1$~K, 
the hump structure in $C(T)$ at $T\sim T^{*}\ll\Delta_{\rm c}$ should 
arise from the 4th band. 
As holes are doped into the 4/7 phase, the chemical potential shifts to 
lower energies inside the 4th band, 
as shown by 
vertical lines 
in the inset of Fig.~\ref{fig:DOS_Tstar}(a). 
In Fig.~\ref{fig:CTST}(a), we plot the energy difference between the top of the 4th band, 
$W_{4}^{\rm top}\equiv E_{4}^{\rm max}({\bf k})$, and the chemical potential $\mu$, 
$W_{4}^{\rm top}-\mu$, 
by vertical lines. 
We see that $W_{4}^{\rm top}-\mu$ is located at the central position of the 
hump of $C(T)$ for each $n_{0}$. 
This confirms that the characteristic energy of the fluid in the 
density-ordered-fluid phase expressed by $W_{4}^{\rm top}-\mu$ 
corresponds to $T^{*}$. 
The filling dependences of $T^{*}$ (open square) and $W_{4}^{\rm top}-\mu$ (open diamond) 
are shown in Fig.~\ref{fig:DOS_Tstar}(b). 

To make a detailed comparison with experiments, we extrapolate the ``charge gap" $\Delta_c$ 
by the least-squares fit of the data for $35~{\rm K}\le V_{\rm cutoff}\le 45$~K 
assuming the form $\sum_{n=0}^{2}a_{n}V_{\rm cutoff}^{n}$~\cite{notegap}, 
which is shown by the dashed line in Fig.~\ref{fig:cgp}(c). 
We also extrapolate the width of the 4th band, 
$W_{4}\equiv E_{4}^{\rm max}({\bf k})-E_{4}^{\rm min}({\bf k})$ 
by the least-squares fit of the data for $35~{\rm K}\le V_{\rm cutoff}\le 80$~K 
(filled circles in Fig.~\ref{fig:cgp}(c)) 
assuming the form $\sum_{n=0}^{3}b_{n}V_{\rm cutoff}^{n}$ and 
the result is represented by the dashed line in Fig.~\ref{fig:cgp}(c). 
Since the ``charge gap" of the 4/7 phase is expected to be 
$\Delta_{\rm c}\sim 1$~K~\cite{Sciver,WI2007}, 
the corresponding $V_{\rm cutoff}$ is estimated to be 22.5~K, 
which is not inconsistent with the chemical potential difference between 
the 2nd and 3rd layers, 16~K~\cite{Whitlock} 
(the gray line in Fig.~\ref{fig:cgp}(c)). 
Then, $W_4$ at $V_{\rm cutoff}=22.5$~K is evaluated to be 174~mK. 
Since $W_{4}(V_{\rm cutoff}=22.5~{\rm K})/W_{4}(V_{\rm cutoff}=60~{\rm K})
\sim 0.43$, the density of states of the 4th band at $V_{\rm cutoff}=22.5$~K 
is inferred to be enhanced $1/0.43$ times more than that at $V_{\rm cutoff}=60$~K, 
as shown in the inset of Fig.~\ref{fig:DOS_Tstar}(a). 
Then, $W_{4}^{\rm top}-\mu$ and $T^{*}$ at $V_{\rm cutoff}=22.5$~K are estimated to be 
$0.43$ times smaller than those at $V_{\rm cutoff}=60$~K, 
which are shown by filled diamonds and filled squares, respectively, 
in Fig.~\ref{fig:DOS_Tstar}(b). 
The slope of the resultant $T^{*}$ is evaluated to be $T^{*}\sim 400\delta$~mK with 
$\delta \equiv 1-n_0$, which is quite consistent with the experimental observation 
$T^{*}\sim 430\delta$~mK~\cite{Fukuyama2007}. 
This analysis shows that 
$T^{*}$ may be regarded as the effective bandwidth 
of the holes by quantum-mechanical zero-point motions in the solid, 
which substantiates the hypothesized idea of 
{\it zero-point vacancy}~\cite{ZPV,Fukuyama2007}.

We note that the filling dependence of the entropy $S(T)$ 
for $T^{*}<T\ll T_{\rm c}$ (for example, $S(T=100~{\rm mK})$, not shown) shows nearly 
the same $n_0$ dependence as 
$S=-n_0{\rm ln}n_0-(1-n_0){\rm ln}(1-n_0)$, 
as experimentally observed~\cite{Fukuyama2007}. 
This implies that the entropy for distributing $N\delta$ holes in the $N$-site system 
can be accounted for by the hole-pockets contribution in the density-ordered-fluid phase.

We note that a broad shoulder structure of $C(T)$ in the uniform fluid phase 
for $n_0=0.3$ (not shown) evolves into the $C(T^{*})$ hump as $n_0$ increases, 
which finally shrinks toward $n_0=1$ as in Fig.~\ref{fig:CTST}(a). 
A similar $C(T)$ evolution was observed in the layered $^3$He system 
on the two $^4$He layers adsorbed on graphite 
by Neumann et al.~\cite{saunders} 
when the $^3$He density $n$ increases and approaches $n_{\rm c}=9.9$~nm$^{-2}$. 
Since $C(T)$ has common features in both systems~\cite{saunders,Fukuyama2007}, 
it is natural to interpret the intervening phase observed for $n_{\rm I}<n<n_{\rm c}$ 
with $n_{\rm I}=9.2{\rm nm}^{-2}$ in ref.~\citen{saunders} as a density-ordered fluid 
stabilized near the density-ordered solid at $n=n_{\rm c}$. 
This offers a clear and alternative interpretation of the sharp transition or crossover 
around $n=9.2$~nm$^{-2}$ in ref.~\citen{saunders}.

The density-ordered fluid whose ground-state energy is lower 
than that of the uniform fluid, $E_{\rm DOF}(n_0)<E_{\rm fluid}^{\rm uni}(n_0)$, 
is confirmed at least up to 7\% hole doping for $V_{\rm cutoff}=60$~K.  
The poor convergence of the MF solution upon 
further doping prevents us from determining the accurate phase boundary. 

Here, we also stress 
an alternative possibility, namely, the emergence of a uniform fluid phase 
with small Fermi surfaces when holes are doped into the density-ordered phase, 
even when 
fluctuations beyond the MF theory destroy the density order in the 
absence of spin order. 
The Fermi surface is defined by 
poles of the single-particle Green function $G({\bf k},\omega)$ 
at a frequency $\omega=0$~\cite{memo_w}. 
While Re$G$ changes its sign across a pole through $\pm\infty$, 
Re$G$ can also change the sign across a $zero$ defined by $G=0$. 
In the solid phase, only the zero surface exists in the $\bf k$ space at $\omega =0$ 
while only the poles exist at $\omega=0$ for heavily doped uniform fluids. 
When holes are slightly doped, the reconstruction of $G({\bf k},\omega)$ yields 
the interference between the zeros and the poles, 
which creates the resultant Fermi surface 
with the coexistence of zeros and poles. 
Since the interference has a significant $\bf k$ dependence at $\omega =0$, 
small Fermi pockets may appear after the truncation of the large Fermi surface~\cite{sakai} 
even when the density order 
is destroyed upon hole doping in the absence of the spin order. 
It is remarkable that not only doped Mott insulators but also doped density-order phases 
show such differentiation as observed in the 2D $^3$He system. 
This differentiation in ${\bf k}$ space can also be the origin of 
the peak (small cusp) at $T\sim J$ and the hump (peak) at $T\sim T^{*}$ in $C(T)$ 
for $n_0<1$~\cite{Fukuyama2007} ($n<n_{\rm c}$~\cite{saunders}), 
since the former and latter are attributed to 
the contributions from the truncated and remaining parts of the Fermi surface 
in $\bf k$ space, respectively. 
A multiscale measurement of $C(T)$ ranging from 1~mK to 1~K 
is desired to resolve this issue in layered $^3$He systems.

\section*{Acknowledgment}

We thank Hiroshi Fukuyama for supplying us with experimental data 
and T. Takagi for showing us his PIMC data 
prior to publication with enlightening discussions on their analyses.
This work is supported by Grants-in-Aid for Scientific
Research on Priority Areas under grant numbers
17071003, 16076212 and 18740191 from MEXT, Japan.



\begin{thebibliography}{99}
\bibitem{memo_dE} 
Similar temperature and doping dependences of specific heat 
appear for $\Delta E_1=-3.0$~K, $\Delta E_2=-1.5$~K and $\Delta E_3=0.0$~K, 
since nearly the same structure of the DOS around the top of the 4th band is realized 
in the MF framework, which is responsible for the low-temperature part of specific heat.
\bibitem{notegap} Since the ``charge gap" shows a linear $V_{\rm cutoff}$ dependence 
for large $V_{\rm cutoff}$, we used small-$V_{\rm cutoff}$ data for the least-squares fit.
\end{thebibliography}
\end{document}